# Hydrodynamic mobility of confined polymeric particles, vesicles and cancer cells in a square microchannel

Shamim M. Ahmmed[1+], Naureen S. Suteria[1+], Valeria Garbin[2] and Siva A. Vanapalli[1]

[1]Chemical Engineering, Texas Tech University, Lubbock, Texas 79409, USA

[2]Department of Chemical Engineering, Imperial College London, London SW7 2AZ, UK

+ Equal contributing author

* Corresponding author: siva.vanapalli@ttu.edu

**Abstract**

The transport of deformable objects including polymer particles, vesicles and cells, has been a subject of interest for several decades where the majority of experimental and theoretical studies have been focused on circular tubes. Due to advances in microfluidics, there is a need to study the transport of individual deformable particles in rectangular microchannels where corner flows can be important. In this study, we report measurements of hydrodynamic mobility of confined polymeric particles, vesicles and cancer cells in a linear microchannel with square cross-section. Our operating conditions are such that the mobility is measured as a function of geometric confinement over the range $0.3 < \lambda < 1.5$ and at specified particle Reynolds numbers that are within $0.1 < \text{Re}_p < 2.5$. The experimental mobility data of each of these systems is compared with the circular-tube theory of Hestroni, Haber and Wacholder (J. of Fluid Mech., 1970) with modifications made for a square cross-section. For polymeric particles, we find that the mobility data agrees well over a large confinement range with the theory but under predicts for vesicles. The mobility of vesicles is higher in a square channel than in a circular tube, and does not depend significantly on membrane mechanical properties. The mobility of cancer cells is in good agreement with the theory up to $\lambda \approx 0.8$, after which it deviates. Comparison of the mobility data of the three systems reveals that cancer cells have higher mobility than rigid particles but lower than vesicles, suggesting that the cell membrane frictional properties are in between a solid-like interface and a fluid bilayer. We explain further the differences in the mobility of the three systems by considering their shape deformation and surface flow on the interface. The results of this study may find potential applications in drug delivery and biomedical diagnostics.

## I. Introduction

Understanding the transport of deformable particles, such as vesicles and cells, in moderately or tightly confined conduits is of practical interest in applications ranging from blood rheology[1, 2] to drug delivery[3, 4] to biomedical diagnostics[5-7]. A key parameter of interest in these studies is the hydrodynamic mobility of the deformable particle, $\beta$, which is defined as the ratio of the particle velocity ($U$) to the surrounding mean fluid velocity ($V$). This mobility is expected to be a function of the particle confinement, $\lambda$, defined as the ratio of the particle diameter ($a$) to the hydraulic diameter of the conduit ($D$), particle location in the conduit, as well as flow-based dimensionless parameters such as capillary number ($Ca$) and Reynolds number ($Re$).

Early studies of theory and measurement of particle mobility have focused on cylindrical geometries. Experimental measurements of particle mobility in cylindrical conduits have been reported with drops[8-10], vesicles[11] and red blood cells[12, 13]. Ho and Leal[8] ($\lambda$ = 0.72 – 1.10, $Ca$ = 0.08-0.18) measured the mobility of neutrally buoyant drops in a Newtonian fluid and a viscoelastic fluid for inner to outer fluid viscosity ratios, $K$ = 0.2-2.0. They found that the mobility monotonically decreased as a function of confinement for 0.72 < $\lambda$ < 0.9. For highly confined drops ($\lambda$ > 0.95), the mobility decreased with increasing viscosity ratio for a given capillary number and increased with increasing capillary number for a given viscosity ratio[10]. Olbricht and Leal[9] measured the mobility of buoyant drops, of diameters comparable to the tube diameter, and found that as the density difference between drop fluid and outer fluid increased, the drop velocity decreased and the droplets experienced larger deformation.

There have been parallel efforts to develop theories for predicting the mobility of deformable particles[14-17], as well as solid particles[18-21], in cylindrical tubes. Hetsroni, Haber and Wacholder[22] (henceforth referred to as HHW) derived an analytical model using the method of reflections to determine the settling velocity of a neutrally buoyant, spherical droplet under Stokes flow in a cylindrical tube. Hyman and Skalak[15] considered a train of equally spaced axisymmetric spherical inertia-less droplets as a model for the flow of blood cells in capillaries. They used the stream function approach and adapted the general solution of Wang and Skalak[21] for a train of elastic particles. Martinez and Udell[16] conducted a numerical analysis study using the boundary integral method to derive the velocity of droplets of sizes comparable to the tube diameter. The mobility calculated from the HHW[22] model was shown to be in good agreement with the experimental data reported by Belloul *et al.*[23] in cylindrical tubes at lower confinement regime, $\lambda < 0.7$. For larger drops ($0.7 < \lambda < 1.1$). , numerical simulations by Martinez and Udell[16] showed that they could match the experimental results of Ho and Leal[8].

With respect to the motion of vesicles in tubes, Vitkova *et al.*[11] measured the hydrodynamic mobility and deformation of vesicles in the confinement range of $0.25 < \lambda < 1.1$ and compared the data with the HHW[22] model, assuming the drop viscosity as infinite, *i.e.* they invoked the rigid sphere approximation. Reasonably good agreement was found with the HHW[22] model, prompting them to suggest that there is no momentum transfer across the bilayer for motion of vesicle in tubes. Bruinsma[24] applied the lubrication theory to describe long and closely-fitting vesicle motion in tubes and discussed the rheological regimes. More recently, Barakat *et al.*[25] developed a singular perturbation theory for the motion of an inertia-less vesicle in a tube and found good agreement with the mobility data of Vitkova *et al.*[11].

The motion of red blood cells (RBCs) in cylindrical tubes has been extensively studied[12, 13, 26-32] in the context of blood rheology, but the number of mobility measurements are much less. Albrecht *et al.*[13] measured the mobility (they referred it as overvelocity) of RBCs in a glass capillary at different hematocrits (0.1- 0.6) and varying capillary diameter (3.3 to 11 µm). They compared the mobility of RBCs with the mobility of model particles measured in a circular tube by Lee *et al.*[31] and Sutera *et al.*[33]. They found that RBCs have higher mobility than the model particles. Halpern *et al.*[32] and Secomb *et al.*[27] also studied the mobility of RBCs in circular tube through numerical analysis. We did not find any studies related to measurement of mobility of cancer cells in tubes.

With advances in microfluidics, there is currently significant interest in studying the transport of particles in rectangular microchannels[34, 35]. In contrast to the numerous investigations in tubes discussed above, relatively few studies exist on studying particle mobility in microchannels. Mietke *et al.*[36] experimentally measured the velocity of rigid polystyrene (PS) beads in a square channel ($Re \approx 0.1$, $\lambda = 0.55 \sim 1$) and good agreement was found with the analytical results obtained from the stream-function approach. With respect to cancer cells, there are several studies[37-41] that measured their passage time or velocity in rectangular or square microchannel at $\lambda > 1$, but attempts to determine mobility were not pursued.

It is clear from our survey of literature that there are several gaps in the experimental measurement of mobility of deformable particles and cells in microchannels. First, mobility data on vesicles and cancer cells in microchannel flow do not exist. Second, there has not been a systematic comparison of the mobility of polymeric particles, vesicles and cells. Such a comparative investigation might allow insights drawn from understanding the mobility of simpler model systems to interpret data from cells that are more complex.

To address these gaps, we focused on the measurement of the mobility of vesicles and tumor cells in a microchannel with square cross-section, supplemented by studies with rigid and elastic spheres. Our measurement regime for particle confinement is $0.3 < \lambda < 1.5$, and the flow conditions are such that particle Reynolds numbers is $0.1 < Re_p < 2.5$. Similar to Vitkova *et al.*[11], we compare the results of our data with the HHW model. Our results show that simple modifications of the HHW model for cylindrical conduits can successfully capture the mobility of rigid and elastic spheres, even when $\lambda \rightarrow 1$. For vesicles and cells, we find that the mobility of cancer cells is larger than that of rigid/elastic spheres but lower than vesicles. We explain our results by considering shape deformation and surface mobility at the interface in the three systems.

## II. Materials and Methods
### A. Samples
***Polymeric particles:*** Polystyrene beads (Polyscience Inc., USA) and polydimethylsiloxane (PDMS, Dow Corning) particles[41] were used as a model for rigid and elastic spherical particles, respectively. The PDMS particles were synthesized using a 30:1 ratio of base to curing agent. 1 mL of the PDMS mixture was added to 10 mL of 3 wt% Tween 20. The particle emulsion was created by mixing the solution for 2 minutes with a vortex mixer and curing it overnight at 70 °C. The particle solution was then filtered with a 30 μm filter (CellTrics, Germany) to obtain particles with a maximum diameter of 30 μm. Two different suspending phases were used: deionized (DI) water and 11 wt.% polyethylene glycol (PEG20000, Fluka Analytical) in phosphate buffered saline (PBS) with viscosities $\mu_o$ = 1 and 10 mPa.s, respectively. The particle solutions were prepared to a final concentration of 5 x 10$^5$ particles/mL. Polystyrene particles had a mean diameter of 15.13 μm ± 6%.

*Vesicles:* Giant unilamellar vesicles (GUVs) were prepared using standard electroformation protocols[42]. The lipids used were 1,2-dioleoyl-*sn*-glycero-3-phosphocholine (DOPC) and 1-stearoyl-2-oleoyl-*sn*-glycero-3-phospho- choline (SOPC). They were diluted with chloroform to achieve a concentration of 1 mg/mL. Vesicles were formed using 2 lipid solutions: DOPC and a 1:1 molar ratio of SOPC and cholesterol (here onwards referred to as SOPC:Chol). An electroformation chamber was created using two indium tin oxide glass slides (15-25 Ω/sq) and a 1 mm thick PDMS spacer. A thin film was created in the chamber by dispensing 10 μL of 1 mg/mL lipid solution onto one of the glass slides and then drying it in a vacuum desiccator for 30 minutes. In the closed chamber, the lipids were hydrated with a 0.11 M sucrose solution. The chamber was then connected to a waveform generator and AC voltage was applied at 2.6 V and 10 Hz for 3 hours and then 4.4 V and 4 Hz for 45 minutes. After formation, the vesicles solution was diluted with 0.12 M glucose. The lipids were purchased from Avanti Polar Lipids and the remaining materials from Sigma Aldrich.

The membrane properties of these two vesicle systems are listed in Table 1, which were obtained from the literature[43-45], with vesicles made from SOPC:Chol having higher values of bending modulus ($\kappa_b$), stretching elasticity ($K_s$) and lysis tension ($\sigma_c$) than DOPC vesicles. Two dimensionless parameters that dictate the influence of thermal fluctuations in the lipid bilayer on the vesicle shape are excess area, $\Delta$, and reduced volume, $v$. Here $\Delta = \frac{S}{4\pi R^2} - 1$, where $S$ is the measured surface area of the vesicle and $R$ is the radius of a sphere with the same measured volume of the vesicle. The reduced volume is the ratio of the actual volume of the vesicle to the volume of a sphere with the same surface area, which can be shown to be related to $\Delta$, as $v = (1 + \Delta)^{-3/2}$.

Garbin and coworkers[46] measured the excess area of these two systems. They measured the surface area and volume of the vesicles by flowing them in cylindrical capillaries and operating at conditions such that the vesicle membrane is flat and ironing out the thermal fluctuations but it is not stretched. These operating conditions correspond to $Ca_b = \mu_0 V R^2 / \kappa_b > 1$ and $Ca_K = \mu_0 V / K_s < 10^{-3}$, where $Ca_b$ and $Ca_K$ are the two capillary numbers based on bending modulus and stretching elasticity, respectively. For the lipid systems used in this study, they measured average excess area values of 0.13 and 0.17 for DOPC and SOPC:Chol, respectively[46]. From these $\Delta$ values, we calculated the mean values for $\nu$ which are 0.83 and 0.79 for DOPC and SOPC:Chol, respectively. Note that $\nu = 1$ corresponds to a fully inflated spherical vesicle.

**Table 1: Membrane properties of the vesicles used in the study. The data was obtained from refs. 43 – 45.**

| Parameter | Symbol | Units | DOPC | SOPC:Chol |
|---|---|---|---|---|
| Bending Modulus | $\kappa_b$ | J | $1.08 \times 10^{-19}$ | $2.46 \times 10^{-19}$ |
| Stretching Elasticity | $K_s$ | mN/m | 310 | 1985 |
| Lysis Tension | $\sigma_c$ | mN/m | 9.92 | 25.805 |

*Cancer cells:* In this study, breast cancer cell lines MCF7 and MDA-MB231 and lung cancer cell lines H1437 and H1299 were used. MCF7 was obtained from Dr. Lauren Gollahon (TTU, Department of Biological Sciences), MDA-MB231 was purchased from ATCC (ATCC® HTB-26™) and H1437 and H1299 were obtained from Dr. Sam Hanash (MD Anderson Cancer Center, The University of Texas). MCF7 and MDA-MB231 were cultured in Dulbecco's modified eagle medium (DMEM) supplemented with 10% fetal bovine serum (FBS), 1% Penicillin/Streptomycin (Gibco 15140-148) and 1% sodium pyruvate (Gibco). H1299 and H1437 were cultured in RMPI 1640 containing 10% FBS and 1% Penicillin/Streptomycin. All cells were cultured in an incubator at 37°C in a 5% $CO_2$ environment. Confluent cells were harvested for experiment using Trypsin/EDTA (0.25%, Gibco). All the experiments were completed within 30 minutes of being harvested. Trypan blue, at a final concentration of 10% v/v, was added to suspending phase to identify dead cells entering to the channel and to have better contrast between cells and the surrounding fluid phase. The cell concentration used in the mobility experiments was 5-6 x $10^5$ cells/mL.

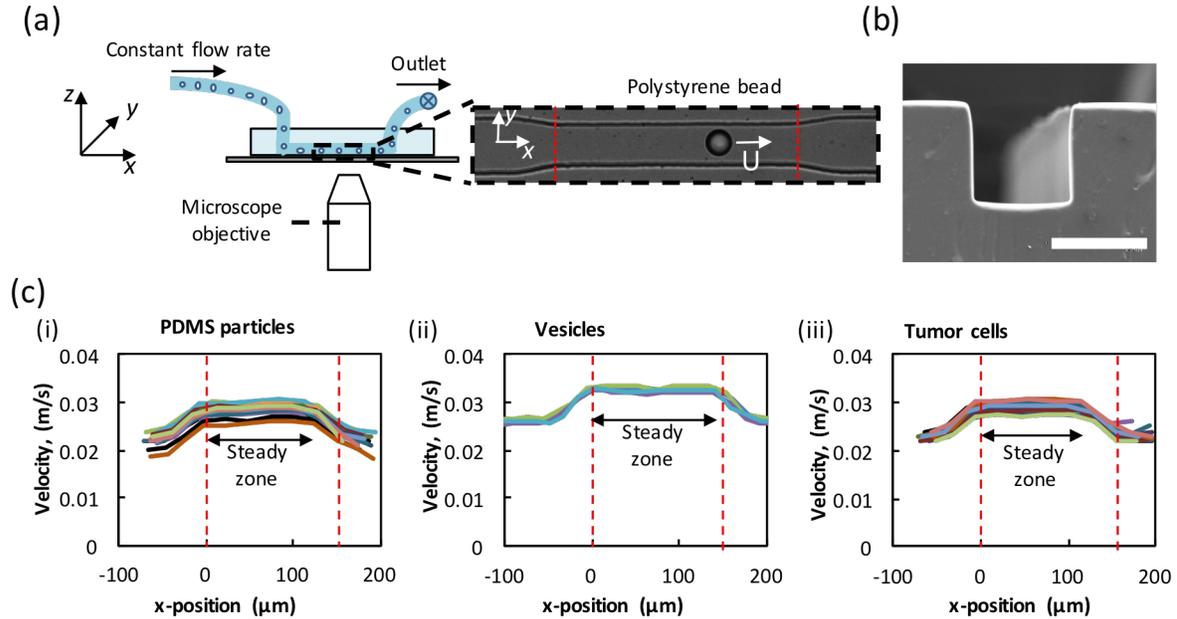

**Figure 1: High throughput particle mobility measurements in microchannels.** (a) Schematic of the experimental setup showing the microfluidic device, microscope objective and the particle flow path through the microchannel. Inset shows the bright field image of the tapered entrance and the straight channel with a polystyrene bead traveling at a velocity $U$. The dashed vertical lines in red indicate the linear section of the microfluidic geometry. The straight channel has a height, width and length of 25±0.4, 25.3±0.4 and 150 µm, respectively. (b) Scanning electron micrograph image showing the square channel cross-section. The scale bar is 25 µm. (c) The instantaneous velocity of i) PDMS particles, ii) vesicles, and iii) tumor (MCF7) cells along the x-position of the channel length. For all three systems, the velocity is plotted for the confinement range of $0.75 < \lambda < 0.80$. The region between the dashed vertical lines in red indicates the linear portion of the microfluidic geometry.

### B. Microfluidic channel fabrication

Linear microchannels of length 150 µm and a square cross-section (measured width is 25±0.4 and height is 25.3±0.4 µm), as shown in Figs. 1a, b, were fabricated using standard soft lithography techniques[47]. The PDMS replicas was cut, peeled, and 1 mm holes were created using biopsy punchers (Miltex, Japan) for connecting the tubing. They were subsequently cleaned using isopropanol and bonded to cover glass (No. 2, Fisher Scientific) using corona treatment[48] for 2 minutes (BD-20AC Laboratory corona treater, Electro-technic products). After bonding, the devices were heated at 70˚C for 4 minutes to have a permanent seal. The channels were filled with 4 wt.% bovine serum albumin (BSA, Sigma Aldrich) in phosphate-buffered saline solution (Gibco) for at least 1 hour at 37˚C to reduce non-specific adhesion of cells to the walls. The sealed devices were used within two days of bonding.

### C. Experimental protocol

The experimental setup for mobility measurement consisted of a microfluidic device, a syringe pump and a microscope connected with a high-speed camera. A simplified schematic of the experimental setup is shown in Fig. 1a. A syringe pump (PHD 2000, Harvard Apparatus, Massachusetts, USA) was used to drive fluid from a 100 μL Hamilton gastight syringe to the microfluidic device through 0.02" inner diameter Tygon tubing (Cole Parmer) and 20-gauge hollow blunt pin (Instech, USA). A constant flow rate of 100 μL/hr was used for the particle experiments and 50 μL/hr was used for the vesicle and cell experiments.

Bright-field imaging was used to record particle and cell passage through the test section of the channel using a combination of an inverted microscope (Nikon eclipse TiU) and high-speed CMOS camera (Phantom v710 12-bit, Vision Research). The region of interest (ROI) included the area of test section and three channel widths before and after the test section. The ROI was recorded with a reduced resolution of 448 x 80 pixels at a frame rate of 2000 fps using 30x magnification. The microscope objective was focused approximately on the midplane of the linear channel. The effective pixel size for this optical setup is 0.64 μm and the depth of focus is ~2.5 μm. Implementation of Köhler illumination combined with a high-power halogen bulb enabled us to use 1 μs exposure time to record blur-free motion of cells traveling at typically ~3.0 cm/s. The image based auto trigger (IBAT) feature of the Phantom v710 camera was used to save only those frames when a cell or particle passes through the test section, which reduces the number of images that needs to be analyzed drastically (about 15 - 20 frames per particle or cell). Imaging of vesicles was performed in the phase contrast mode on an inverted microscope (IX71, Olympus Inc.) connected with a CMOS camera (Phantom v310 12-bit, Vision Research). A frame rate of 5000 fps was used to record images at 32x magnification and 50 μs of exposure time. The effective pixel size for this optical setup is 0.625 μm.

### D. Image processing

A custom written MATLAB routine was used for the automated image processing and data analysis of particles and cells. For vesicle experiments, analysis was performed manually using ImageJ software (NIH) due to the low contrast between the vesicles and the surrounding fluid. For automated analysis, images were segmented using different filters to enhance contrast, subtract background and identify the presence of an object, in this case a particle or a cell. After segmentation, each object was given an identification (ID) number when it appears in the ROI for the first time. The object's projected area (from top view), centroid location, perimeter and frame number were recorded against that ID number for all the frames that the object takes to pass through the ROI. The object diameter was calculated from the cross-sectional area assuming that the cross-sectional area represents the maximum cross-section of a sphere (max error in size measurement <3%). MATLAB's built in function 'regionprops' in the image processing toolbox was used to obtain the area, perimeter and centroid location from the segmented image.

Image frames where multiple particles or cells occupied the ROI were discounted since it is possible that their mobility is affected due to hydrodynamic interactions. The degree to which the object was off-centered (in *xy*- plane) in the channel was measured from the *y*-coordinate of the centroid and the mobility data was discarded if the centroid deviated by more than 2.0% from the center. We also measured the shape deformation of each particle inside the test section of the channel and only picked the particles which are spherical in shape using a threshold deformation index value (discussed in section IV A).

### E. Flow conditions

Table 2 reports the particle-based Reynolds number and capillary numbers corresponding to the flow conditions used in this study. The particle-based Re[49] is defined as $Re_p = \rho_0 V_{max} a^2 / \mu_0 D$, where $\rho_0$ is the density of the suspending fluid and $V_{max}$ is the maximum fluid velocity in the channel. Because of the polydispersity present in the systems being studied, here we use the mean particle diameter *a* to calculate $Re_p$ which ranged from 0.09 to 2.47. This range indicates that particles, vesicles and cells all exhibited finite inertia while passing through the microchannel. For PS and PDMS particles, the capillary number is defined as, $Ca = \mu_0 V / aG'$, where $G'$ is the elastic modulus of the particle (30:1 PDMS $G' = 95$ kPa[41]; PS $G' = 3.25$ GPa[50]). For the PS and PDMS particles, Ca <<1, suggesting that the particles did not deform while flowing through the test channel. For cells, the capillary number is defined as $Ca = \mu_0 V / \gamma$, where $\gamma$ is the membrane tension of the cell (0.01-1 mN/m[51, 52]). The Ca values ranged from 0.017-2, suggesting that the cells did not deform significantly in the microchannel during flow.

**Table 2: Experimental parameters and regime of operation for mobility measurement.**

|  | Samples | $Re_p$ | $Ca$ | $Ca_b$ | $Ca_K$ |
|---|---|---|---|---|---|
| Rigid particles | PS[1] | 0.920 | $1.63 \times 10^{-9}$ | - | - |
|  | PS[2] | 0.092 | $1.63 \times 10^{-8}$ | - | - |
|  | PDMS[1] | 1.333 | $5.59 \times 10^{-5}$ | - | - |
|  | PDMS[2] | 0.133 | $5.59 \times 10^{-4}$ | - | - |
| Vesicles | DOPC | 2.471 | - | $3.6 \times 10^4$ | $1.1 \times 10^{-4}$ |
|  | SOPC: Chol | 2.471 | - | $2.1 \times 10^4$ | $1.5 \times 10^{-5}$ |
| Cells | MCF7 | 0.863 | 0.02-2.0 | - | - |
|  | MB231 | 0.863 | 0.02-2.0 | - | - |
|  | H1299 | 1.029 | 0.017-1.7 | - | - |
|  | H1437 | 1.029 | 0.017-1.7 | - | - |

[1] $\mu_0 = 1$ mPa.s, [2] $\mu_0 = 10$ mPa.s

## III. Theoretical model for particle mobility

In this study, we compare the measured mobility of particles, vesicles and cells with predictions from the HHW model. HHW determined the mobility of a neutrally buoyant, small spherical droplet in a circular tube under Stokes flow condition using the method of reflections. Their result is:

$$\beta = 2 \left[ 1 - \left(\frac{b}{R}\right)^2 - \left(\frac{2K}{3K+2}\right) \lambda^2 \right] + O(\lambda^3) \qquad (1)$$

where *b/R* is the offset from the axis of the tube.

Here we chose the HHW model to compare with our experimental data for several reasons. First, unlike HHW, other studies such as those of Hyman and Skalak[15], Martinez and Udell[16], and Wang and Skalak[21] requires numerical analysis to predict mobility. Moreover, the analysis by Martinez and Udell[16] and Wang and Skalak[21] show that their computed hydrodynamic mobility of droplets in circular tubes is in good agreement with the small deformation model of HHW. Second, Murata[19] derived a mobility model for an incompressible, neutrally buoyant, spherical, homogeneous, elastic particle in circular tube under Stokes flow condition, and the HHW result matches that of Murata, when the condition for an elastic particle, $K\rightarrow\infty$, is imposed in the HHW solution. Third, Belloul et al.[23] showed that the HHW result agrees reasonably well with their measured droplet mobility in cylindrical tubes for $\lambda < 0.7$.

To apply the HHW model to our data from square microchannels, we modified Eqn. (1) in the following way. In a circular tube, the ratio of max-to-mean fluid velocity for Poiseuille flow is 2, whereas in square microchannel it is 2.096[53]. We replace the tube diameter with the hydraulic diameter of the square channel. With these adjustments Eqn. (1) becomes,

$$\beta = 2.096 \left[1 - \left(\frac{b}{R_h}\right)^2 - \left(\frac{2K}{3K+2}\right)\lambda^2\right] + O(\lambda^3) \qquad (2)$$

For a solid particle, setting $K\rightarrow\infty$, we obtain

$$\beta = 2.096 \left[1 - \left(\frac{b}{R_h}\right)^2 - \frac{2}{3}\lambda^2\right] + O(\lambda^3) \qquad (3)$$

In this study, Eqns. (2) and (3) were used to compare the mobility of rigid spheres, elastic particles, vesicles and cells.

## IV. Results

### A. Quantification of particle mobility in channel flow

As discussed in the Introduction, the hydrodynamic mobility of a particle is defined as the ratio of the particle's steady velocity to the mean velocity of the surrounding fluid. In this section, we discuss three important factors that can affect the determination of the particle mobility: (i) the three-dimensional position of the particle in the square conduit (ii) the slight non-spherical shape of the particle and (iii) the region in the linear channel where the particle achieves steady velocity.

The three-dimensional location of a particle in the square conduit can influence its mobility with particles close to the wall moving less than those in the center. In our experiment, we are only able to determine the centroid of the particle in the x-y plane (see Fig. 1a) and do not have any control over the z-location of the particle. All the mobility data shown corresponds to the particle centroid (in the x-y plane) being within 1% of the conduit axis. Therefore, most of the uncertainty in determining mobility at a given particle confinement is due to variability in the z-position.

We also considered the influence of the particle shape on its mobility since both cells and vesicles can be non-spherical prior to entering the linear channel or undergo deformation in the channel and become non-spherical. In addition, it is important to consider shape effects, since the HHW model is applicable only for spherical objects. To characterize the object's shape and only consider for mobility analysis those particles (and vesicles and cells) that are spherical in shape prior to entering the linear channel, we defined a deformation index, DI, as[54]

$$DI = 1 - \frac{2\sqrt{\pi A}}{p} \qquad (4)$$

In Eqn. 4, *A* is the projected area of the object as seen in the microscope image and *p* is its perimeter. If the object is completely circular in two-dimensional view, the DI will be zero, which implies a perfect circle. As the object deviates from a perfect circle, the DI will increase accordingly. We measured particle DI prior to entering the linear channel, where the instantaneous velocity starts to increase (referred to as Initial DI).

Fig. 2 shows the Initial DI for the all the systems studied as a function of confinement, where we have only plotted those data points where DI < 0.03. We note that some vesicles and cancer cells were found to have DI > 0.03 which were not included in the mobility analysis. The Initial DIs of rigid PS particles were found to be very small, indicating that as expected these particles enter the channel as spheres. There is some scatter in the DI data even for rigid spherical PS beads. This might be because of variation in the pixel intensity near the edge of the particle due to its rotation in flow or possible variation in z-location of particles. Based on these polystyrene beads results, we use 0.03 as a cut-off value of Initial DI, to assess whether the PDMS particles, vesicles and cells entering the channel are spherical. We analyzed only those vesicles and cancer cells having an initial DI ≤ 0.03 (Fig. 2c, d). Since we measured the initial DI for all the systems before entering into the constricted channel where all the deformable objects are unconfined, we do not see any effect of confinement on the initial DI.

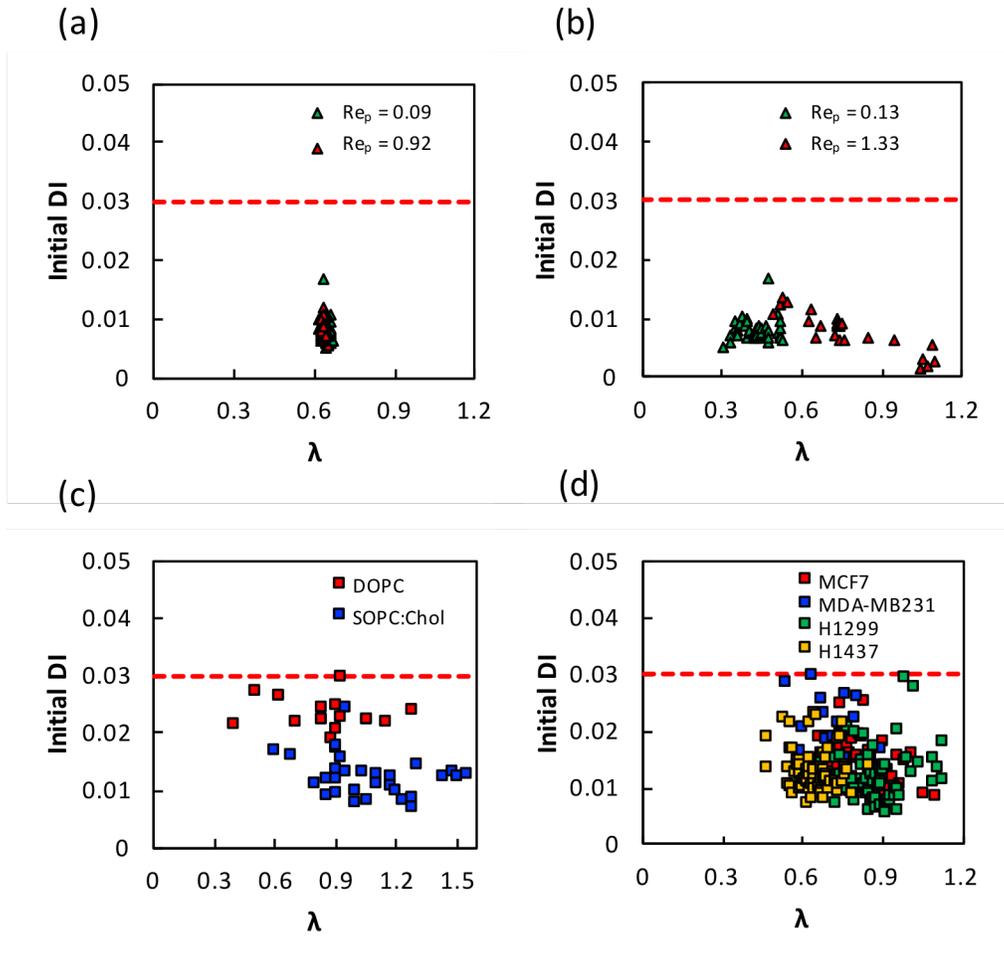

**Figure 2: Deformation index of polymeric particles, vesicles and cancer cells.** The deformation index is plotted as a function of confinement for (a) polystrene beads, (b) PDMS particles, (c) vesicles and (d) cancer cells. The red dashed lines represent a DI = 0.03.

Finally, the region in the linear channel where the particle achieves steady velocity needs to be identified for accurate measurement of mobility. To determine this steady zone, we determined the instantaneous particle velocity as it travels in the microfluidic geometry. Fig. 1c shows the instantaneous velocities for (i) PDMS particles, (ii) vesicles and (iii) cells. In all cases, we find that as the particles enter the tapered entrance into the linear channel, their velocities begin to increase and reaches a maximum at the beginning of the linear channel. To identify the region in the linear channel where the particle velocity is steady, we chose particles in the confinement range of $0.75 \leq \lambda \leq 0.80$ and computed their velocity profiles as shown in Fig. 1c. We find that the velocities remain steady for at least the first ~90% of the channel length before beginning to decrease as they exit into the larger exit channel. We compute the mean of the instantaneous velocities in this steady region and take it as *U* for mobility calculation. We find the maximum variation in particle velocity measurement in this region is < 1 %.

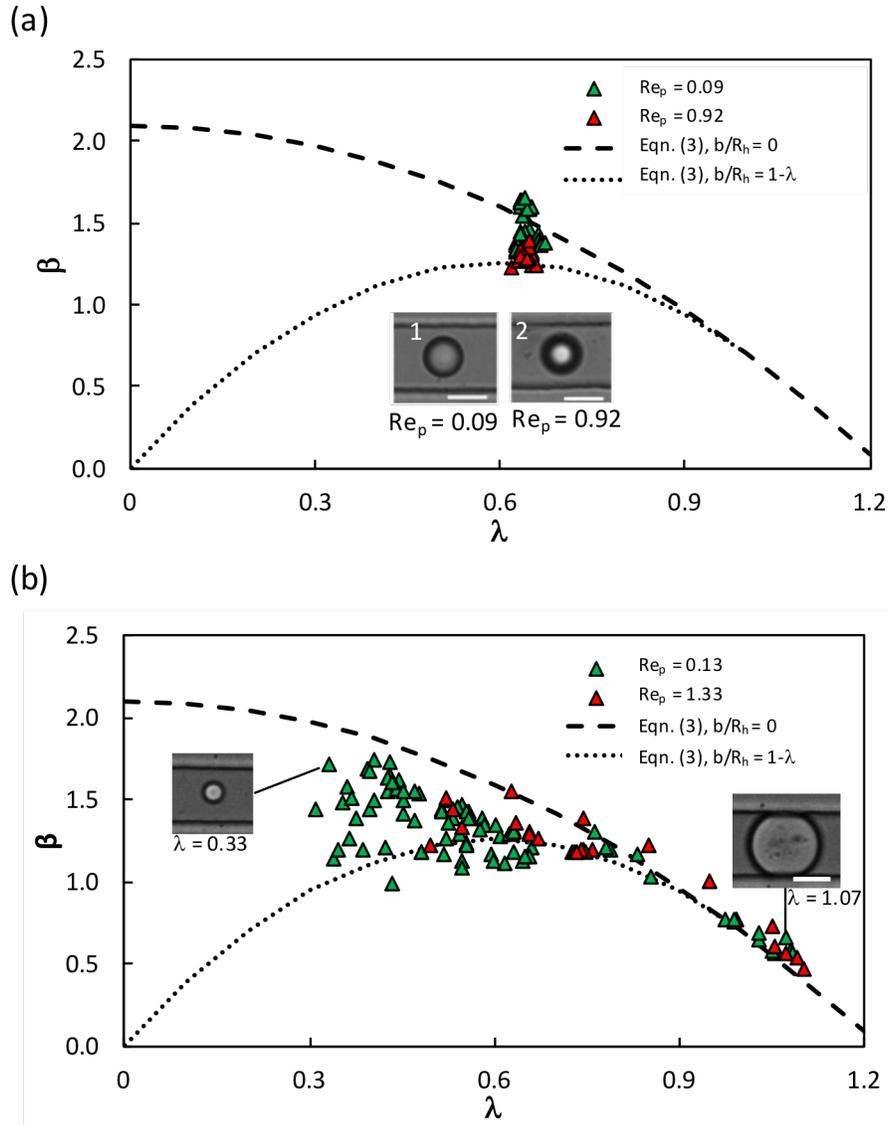

**Figure 3: Mobility of rigid and elastic particles.** The mobility of (a) PS beads and (b) PDMS particles is plotted as a function of confinement. The red and green triangles represent continuous phase viscosities corresponding to $Re_p$= 0.92 and $Re_p$=0.09 for PS beads and $Re_p$= 1.33 and $Re_p$=0.13 for PDMS particle, respectively. Two mobility model curves using Eqn. (3) are shown where the dashed line represents a particle in the center of the channel ($b/R_h = 0$) and a particle with the maximum off-centeredness ($b/R_h = 1- \lambda$). The inset in (a) shows a particle in focus that is in the center of the channel (1) and a particle out of focus that is at the roof of the channel (2). The insets in (b) show experimental images of the PDMS particles at different confinements. The scale bar in all the images is 15 μm.

### B. Mobility of polymeric particles

In this section, we present the mobility of (i) rigid polystyrene particles and (ii) elastic PDMS particles and compare them with the HHW model, Eqn. (3), to determine if the data is in good agreement with the model and over what confinement range the model is applicable.

*Rigid PS spheres:* Fig. 3a shows the hydrodynamic mobility of monodisperse rigid PS particles as a function of confinement at two values of $Re_p$ corresponding to the two different suspending fluid viscosities. To compare this data with the HHW model, we plot Eqn. (3) for the case where the particles are on the centerline, i.e. $b/R_h = 0$, and the case where the particles are touching the channel walls, i.e. $b/R_h$ is maximum. Note that for a given $\lambda$, the maximum possible $b/R_h$ is $1 - \lambda$. The reason we plotted these two curves in Fig. 3a is that we filtered our data to only include those particles whose centroid was within 1% of the centerline. Here the scatter could arise due to particles having different z-locations, nevertheless, we find that majority of the data points for the rigid spheres lie within these two bounds calculated from the HHW model.

An interesting observation from Fig. 3a is that particles at $Re_p = 0.92$ seems to have lower mobility than particles with $Re_p = 0.09$. This could be because the particle inertia at $Re_p = 0.92$ is about ten times more than that at $Re_p = 0.09$; the particles at $Re_p = 0.92$ might therefore be subjected to a higher inertial lift force making them move away from the centerline and reducing their mobility. Alternatively, it is also possible that due to the greater density mismatch for particles with $Re_p = 0.92$, they might have lower mobility. To provide evidence that the lower mobility is due to particles moving closer to the wall, in the inset of Fig. 3a, we show representative images of particles with $Re_p = 0.92$ and $0.09$. Indeed, we observe that at $Re_p = 0.92$, the particle appears defocused indicating it is away from the focal plane and therefore located closer to the channel wall.

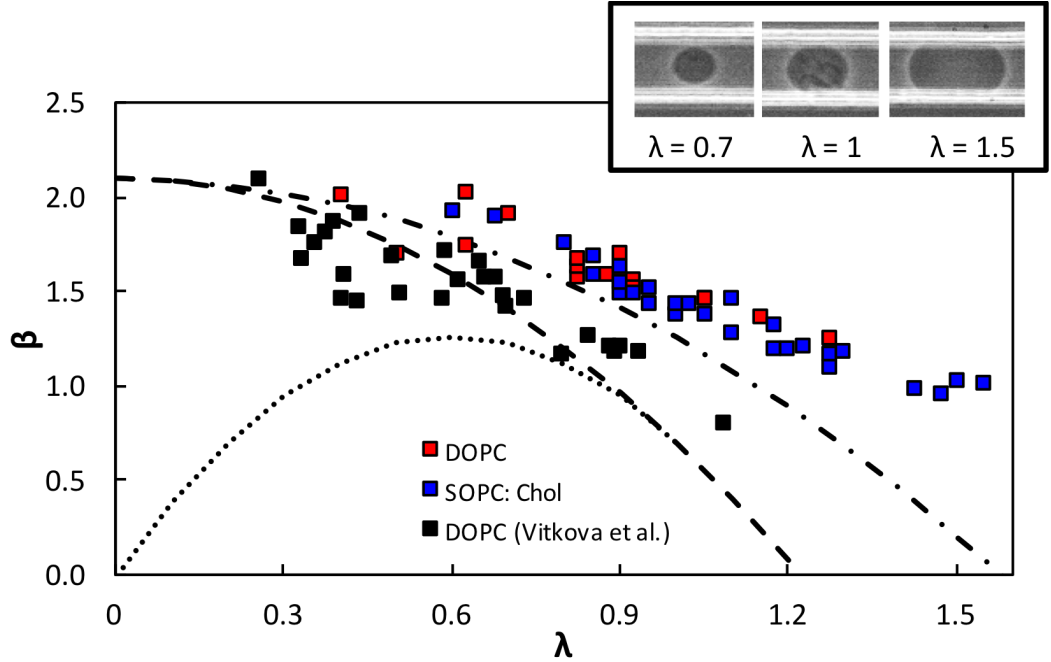

**Figure 4: Mobility of vesicles verses confinement**. DOPC and SOPC:Chol vesicles are the red squares and blue squares, respectively. Vitkova *et al*. data is plotted on top of our experimental data (black square). The dashed line is Eqn. (3) with $b/R_h=0$ and the dotted line is also Eqn. (3) with $b/R_h = 1- \lambda$ for a given $\lambda$. The dot and dash line is Eqn. (2) with K=1. Inset shows experimental images of SOPC:Chol vesicles at different confinements.

*Elastic PDMS particles:* Due to the monodispersity of the PS particles, we were able to determine their mobility at a unique value of confinement $\lambda \approx 0.65$. To evaluate how the mobility of solid particles changes over a wide confinement range, we measured the mobility of polydisperse elastic PDMS particles. In Fig. 3b, as expected we find that the mobility of the PDMS particles decreases with increase in confinement since larger particles occlude more of the conduit space. When comparing the mobility data with the HHW model across the entire confinement range, $0.3 \leq \lambda \leq 1.1$, we find reasonably good agreement. Interestingly, the HHW model is able to predict the mobility up to $\lambda = 1.1$, where PDMS particles are slightly deformed and touching the walls (see the inset in Fig. 3b). This is in contrast to the case of confined droplet motion in cylindrical tubes, where the HHW model could not predict the mobility of droplets with $\lambda \geq 0.7$ because of drop deformation[8, 23]. Another noteworthy observation is that for particles with confinements of $\lambda = 0.40 - 0.65$, we find that some of the data points have mobility less than the lower bound from the HHW model. It is possible that at these low confinements and near-wall condition, the simple modification of the HHW model, Eqn. (3), is not sufficient because the hydrodynamics is more similar to motion of a small particle near a planar wall rather than a large off-centered particle in a square duct.

### C. Mobility of vesicles

In this section, we report the mobility of vesicles formed from two different lipid systems and compare them to the HHW model and data from Vitkova *et al*[11]. In Fig. 4, we plot the mobility of the DOPC and SOPC:Chol vesicles as a function of confinement by considering only those vesicles with initial DI < 0.03. Similar to the polymeric particles, we observed that the vesicle mobility decreased as the confinement increased. Also, even though the mechanical properties of SOPC:Chol were different from that of DOPC (see Table 1), there was not a significant difference in their mobility.

In Fig. 4, we also compare our mobility data with that from the study of Vitkova *et al.*[11] for DOPC vesicles in a circular tube. We find that the vesicle mobility in square channels is higher than that in tubes. Interestingly, the data of Vitkova *et al.* is in good agreement with the HHW model for a solid particle, i.e. $K \rightarrow \infty$. However, in our square channels, the vesicle mobility is higher than the predictions from the HHW model. Vitkova *et al.* explain the good agreement comes from the absence of flow on the vesicle surface due to axisymmetry in the exterior fluid flow. Since the square channel has only planar symmetry, there could be flow on the vesicle surface. To test this hypothesis, we plot the HHW model for droplets with $K = 1$, the case where the interface is completely mobile. We observe a good agreement between our vesicle data and the droplet model suggesting that the vesicle surface might be mobile.

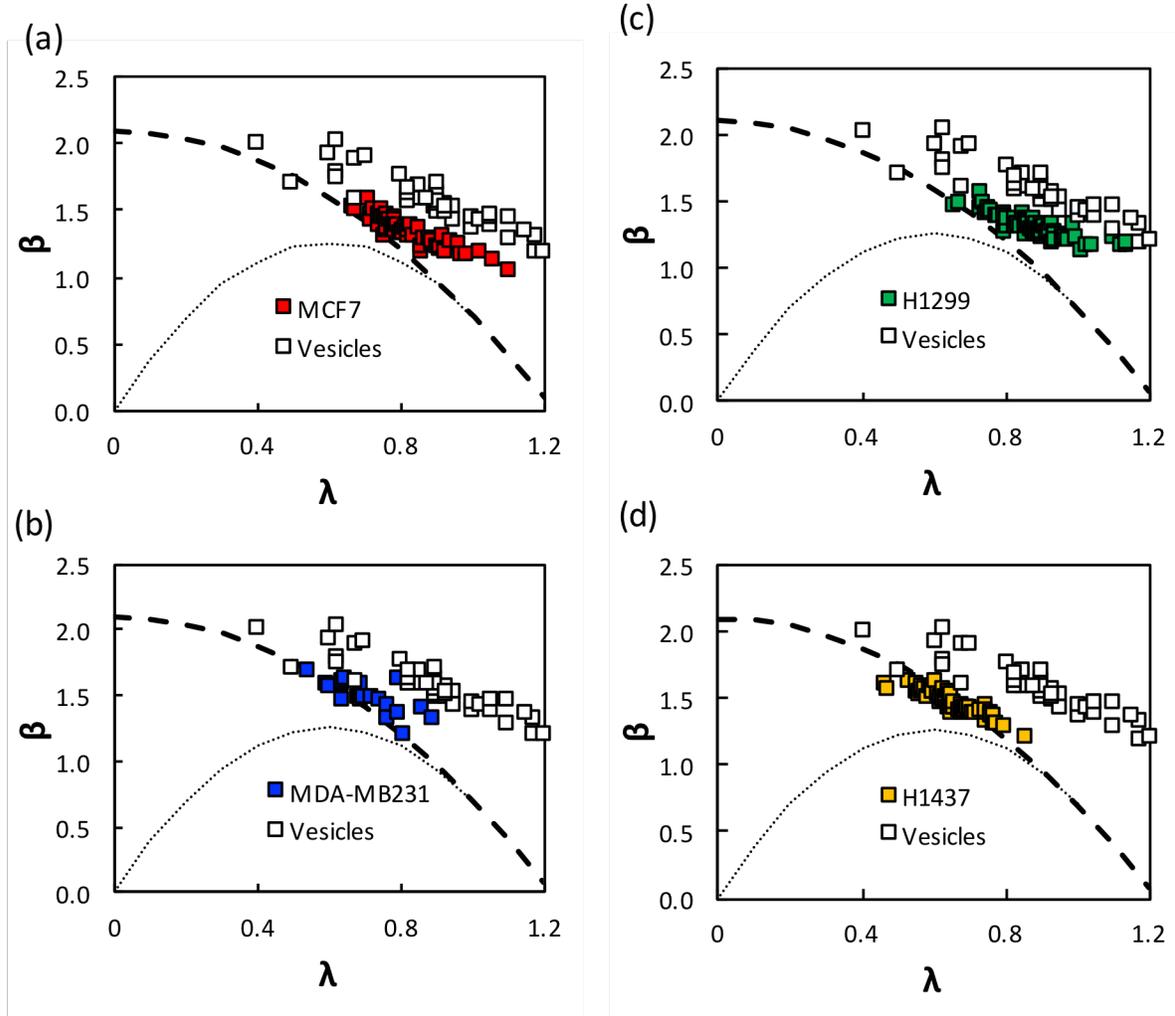

**Figure 5: Mobility of cancer cells.** Mobility versus confinement plot of breast cancer cell line (a) MCF7, (b) MDA-MB231, and lung adenocarcinoma cell line (c) H1299 and (d) H1437. Mobility of vesicles is plotted as white squares for comparison with cells. Also plotted is Eqn. (3) for two conditions of $b/R_h = 0$ (dashed line) and $b/R_h = 1- \lambda$ (dotted line).

### D. Mobility of cancer cells

Next, we measured the mobility of four different cancer cell lines to assess how different is their mobility compared to rigid particles and vesicles. This comparison between model systems and cells is essential to interpret the mobility data of cancer cells, and understand the differences in response of cell and vesicle membrane to hydrodynamic stresses. We considered two breast cancer cell lines, MCF7 and MDA-MB231, with MCF7 being weakly metastatic and MDA-MB231 being highly metastatic[55]. We also considered two lung adenocarcinoma cell lines,

H1299 and H1437, which have features of mesenchymal and epithelial cells, respectively[56]. During metastasis, epithelial cells lose their adherent junctions and switch to a mesenchymal phenotype allowing them to migrate and invade [57, 58]. Selecting these different cell lines also allowed us to examine whether hydrodynamic mobility could be used to elicit differences between cancer cell lines.

Fig. 5a-d shows the mobility results of breast and lung cancer cell lines. Here we have also plotted our vesicle mobility data along with predictions from the HHW model for cases when $b/R_h = 0$ and $b/R_h = 1- \lambda$. Similar to polymeric particles and vesicles, we observe that mobility of cancer cells decreases with increasing confinement. The mobility data for all cancer cells is in good agreement with the HHW model up to $\lambda \approx 0.8$. Interestingly, we find that tumor cells have higher mobility than rigid particles but lower than vesicles, suggesting that the membrane frictional properties are in between a solid-like interface and a fluid bilayer.

Differentiating the cancer cell lines based on mobility was difficult because the mean cell size was not the same in all cell lines. As shown in Fig. 5a, b the confinement for MCF7 and H1299 cells spanned up to $\lambda \approx 1.1$, whereas for MDA-MB231 and H1437 it was limited to $\lambda \approx 0.8$. By comparing the mobility data for $\lambda \leq 0.8$, we observe there are no significant differences in the four cell lines. Here we have used a single microchannel (of cross-sectional area $25 \times 25$ μm$^2$), however, conducting more studies in microchannels of additional cross-sectional areas can produce a wider range of confinement allowing cancer cells of different metastatic capacity to be distinguished based on hydrodynamic mobility.

## V. Discussion

Here, we presented the measurements of mobility for three different systems, namely polymeric particles, vesicles and cells in a square conduit. In all cases, we reported the mobility only for those objects that were spherical prior to entering the linear channel. Then, we compared the experimental measurements with the mobility theory for particle motion in a circular tube with modifications made for square channel. Our key results are: (i) the mobility of rigid and elastic particles are well predicted by the modified HHW model, even at higher confinements (0.7 <

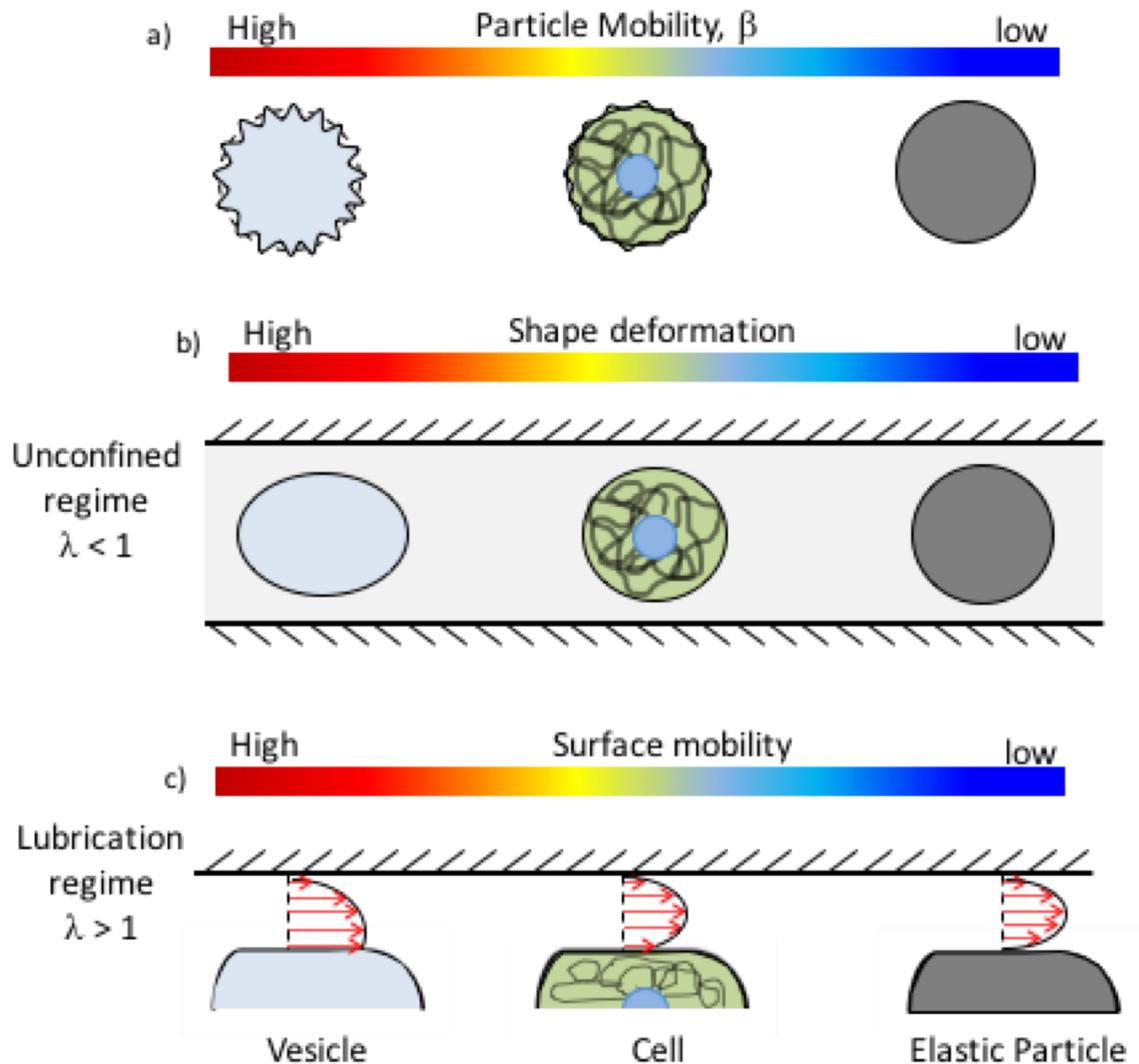

**Figure 6: Possible mechanisms to explain the differences in mobility of vesicles, cells and solid particles.** Variation in (a) particle mobility, (b) shape deformation and (c) surface mobility in the three particle systems studied. The color bar ranges from high (red) to low (blue).

$\lambda<1$), (ii) vesicles move faster in a square channel than in a circular tube and their mechanical properties did not significantly affect their mobility for the conditions studied here (iii) cancer cells have mobility that is lower than vesicles but higher than solid particles. In this section, we discuss the mechanisms that could lead to differences in mobility between the three systems studied.

In Fig. 6a we show the three systems with a color bar indicating the mobility from high to low. When they are unconfined ($\lambda < 1$) as shown in Fig. 6b, vesicles can deform under shear flow, with their deformation being more than cells and rigid particles. This is consistent with the images shown in the inset of Fig. 4, where a vesicle with $\lambda = 0.7$ appears deformed, i.e. major axis 5% larger than minor axis. We did not observe any strong effect of the mechanical properties on the vesicle mobility since our operating regime is such that the bending fluctuations are ironed out ($Ca_b \gg 1$). Although, we could not control the reduced volume in our experiments, we expect that the vesicles with a lower reduced volume will have a higher mobility[25]. In contrast, cells can deform, but much less because their membrane is bound to the cytoskeleton. These deformations effectively reduce their confinement and therefore produce higher hydrodynamic mobility.

Under conditions of stronger confinement *i.e.* $\lambda > 1$, the vesicles and cells can conformally fit to the square cross-section producing thin lubricating films and corner flows. We observe that in this lubricated regime as well, vesicle mobility is higher than the mobility of cells and solid elastic particles. A possible explanation for the higher mobility of vesicles is that their surface is mobile due to the swirling exterior fluid flow created by the presence of corners in a square duct. This surface mobility can admit non-zero velocity gradients on the membrane causing surface flow (see Fig. 6c). In a circular tube, flow around a vesicle in the annulus is axisymmetric that does not permit surface flow on the vesicle membrane[11]. In contrast, motion of a vesicle in a square duct can generate surface flows creating effectively, a slip-like interface and therefore permitting the vesicle to move faster. Indeed, when we assume a fully mobile interface for the vesicle and fit the HHW model for a droplet, the vesicle mobility data is in good agreement with the theory (see Fig. 4). Furthermore, a recent study shows the presence of surface flows on a vesicle anchored to a solid surface and subjected to simple shear flow in a microfluidic device[59].

With regards to cells, their surface mobility might be less than vesicles due to the presence of membrane inclusions such as proteins. The fluidity in the plasma membrane of cells is well established[60] and has been linked to invasion[61] and drug resistance[62] in cancer cells. In our study, we did not find significant differences in the mobility of breast and lung cancer cells for $\lambda < 0.8$, suggesting that in this regime, the frictional properties of the membranes in all the cell lines are similar. More studied need to be pursued over a broader range of confinement to establish the validity of hydrodynamic mobility as a marker for distinguishing cancer cell lines.

## VI. Conclusions

In this study, we presented the mobility results for solid particles, vesicles and tumor cells. For polymeric particles, we found good agreement between our experimental data and a previously established analytical model of HHW. Interestingly, we discovered that the mobility of cells and vesicles was greater than those of the solid particles. In addition, we found that vesicle mobility is higher in a square channel than a circular tube. We explained our observations by considering differences in shape deformation and surface mobility of the three systems studied. In the future, more studies are warranted to confirm the presence of surface flow during vesicle motion in channel flows. The experimental results of mobility reported in this study will perhaps motivate new theoretical and numerical simulations on motion of deformable particles in microchannel flows.

## VII. Acknowledgements

We thank Dr. Lauren Gollahon and Dr. Sam Hanash for providing the cell lines used in this study. We are grateful to Joseph M. Barakat and Professor Eric S. Shaqfeh for useful discussions on vesicles and for sharing their manuscript prior to publication. We thank Angelo Pommella for guidance with the electroformation of vesicles and Md Jasim Uddin for acquiring the SEM image of the microchannel. This work was partially supported by the National Science Foundation (CAREER Grant No. 1150836), CPRIT (Grant No. RP140298) and the Royal Society International Exchange Scheme.